\documentclass{jpsj-suppl}
\usepackage{txfonts} %Please comment out this line unless the txfonts package is availabe in your LaTeX system.

\def\syjm#1#2{{}_{#1}Y_{#2}}
\def\template#1#2#3#4{#1^{(#2)#4}_{#3}}
\def\acoef#1#2{\template{a}{#1}{#2}{}}

\def\m{m_\ps}

\def\twiddle{\lower4pt\hbox{\hskip-0pt{$\widetilde{}$}}}
\def\mth@th{\mathsurround=0pt}
\def\cmapstochar{\mathrel{\rlap{
  \lower0.1pt\hbox{\hskip-1.75pt{$\mapstochar$}}}
  \raise0pt\hbox{\hskip2.5pt{$\twiddle$}}}}
\def\notsimfill{$\mth@th\cmapstochar$}
\def\scroodle#1{\vbox{\ialign{##\crcr\notsimfill\crcr
  \noalign{\kern-4pt\nointerlineskip}
   $\hfil\displaystyle{#1}\hfil$\crcr}}}

\def\ctw{\scroodle{c}{}}

\def\bc#1#2{\left(\begin{smallmatrix} #1 \\ #2 \end{smallmatrix}\right)}

\title{Antimatter, Lorentz Symmetry, and Gravity}

\author{Jay D.\ \textsc{Tasson}}

\inst{St.\ Olaf College, 1520 St.\ Olaf Avenue, Northfield, Minnesota, 55057 USA}

\email{jtasson@carleton.edu}

\recdate{June 28, 2016}

\abst{A brief introduction to the Standard-Model Extension (SME)
approach to testing CPT and Lorentz symmetry
is provided.
Recent proposals for tests
with antimatter
are summarized,
including gravitational and spectroscopic tests.
}

\kword{Lorentz violation, CPT symmetry, Antimatter}

\begin{document}
\maketitle

\newcommand\etal {{\it et al.}}

\newcommand\al{\alpha}
\newcommand\be{\beta}
\newcommand\ga{\gamma}
\newcommand\de{\delta}
\newcommand\ep{\epsilon}
\newcommand\ve{\varepsilon}
\newcommand\ze{\zeta}
\newcommand\et{\eta}
\newcommand\vt{\vartheta}
\newcommand\io{\iota}
\newcommand\ka{\kappa}
\newcommand\la{\lambda}
\newcommand\vpi{\varpi}
\newcommand\rh{\rho}
\newcommand\vr{\varrho}
\newcommand\si{\sigma}
\newcommand\vs{\varsigma}
\newcommand\ta{\tau}
\newcommand\up{\upsilon}
\newcommand\ph{\phi}
\newcommand\vp{\varphi}
\newcommand\ch{\chi}
\newcommand\ps{\psi}
\newcommand\om{\omega}
\newcommand\Ga{\Gamma}
\newcommand\De{\Delta}
\newcommand\Th{\Theta}
\newcommand\La{\Lambda}
\newcommand\Si{\Sigma}
\newcommand\Up{\Upsilon}
\newcommand\Ph{\Phi}
\newcommand\Ps{\Psi}
\newcommand\Om{\Omega}
\newcommand\cA{{\cal A}}
\newcommand\cB{{\cal B}}
\newcommand\cC{{\cal C}}
\newcommand\cE{{\cal E}}
\newcommand\cl{{\cal L}}
\newcommand\cL{{\cal L}}
\newcommand\cO{{\cal O}}
\newcommand\cP{{\cal P}}
\newcommand\cR{{\cal R}}
\newcommand\cV{{\cal V}}
\newcommand\mn{{\mu\nu}}

\newcommand\fr[2]{{{#1} \over {#2}}}
\newcommand\half{{\textstyle{1\over 2}}}
\newcommand\quar{{\textstyle{1\over 4}}}
\newcommand\eigh{{\textstyle{1\over 8}}}
\newcommand\fracn[2]{{\textstyle{{#1}\over {#2}}}}

\newcommand\prt{\partial}

\newcommand\thpr{{these proceedings}}

\newcommand\pt[1]{\phantom{#1}}
\newcommand\ol[1]{\overline{#1}}

\newcommand\vb[2]{e_{#1}^{{\pt{#1}}#2}}
\newcommand\ivb[2]{e^{#1}_{{\pt{#1}}#2}}
\newcommand\uvb[2]{e^{#1#2}}
\newcommand\lvb[2]{e_{#1#2}}
\newcommand\etul[2]{\et^{#1}_{{\pt{#1}}#2}}
\newcommand\hul[2]{h^{#1}_{{\pt{#1}}#2}}

\newcommand\ab{\overline{a}{}}
\newcommand\bb{\overline{b}{}}
\newcommand\cb{\overline{c}{}}
\newcommand\db{\overline{d}{}}
\newcommand\eb{\overline{e}{}}
\newcommand\fb{\overline{f}{}}
\newcommand\gb{\overline{g}{}}
\newcommand\Hb{\overline{H}{}}

\newcommand\kb{\overline{k}{}}
\newcommand\sbar{\overline{s}{}}
\newcommand\kfb{(\overline{k_F})}

\newcommand\psb{\overline{\ps}{}}

\newcommand\mt{m^{\rm T}}
\newcommand\ms{m^{\rm S}}
\newcommand\mb{m^{\rm{B}}}
\newcommand\mbp{m^{\prime\rm{B}}}
\newcommand\qt{q^{\rm T}}

\newcommand\af{(a_{\rm{eff}})}
\newcommand\afp{(a^{\prime{\rm B}}_{\rm{eff}})}
\newcommand\afB{(a^{\rm B}_{\rm{eff}})}
\newcommand\cuB{(c^{\rm B})}
\newcommand\aft{(a^{\rm T}_{\rm{eff}})}
\newcommand\afs{(a^{\rm S}_{\rm{eff}})}

\newcommand\afb{(\ab_{\rm{eff}})}
\newcommand\afbb{(\ab^{\rm B}_{\rm{eff}})}
\newcommand\afbnp{\ab_{\rm{eff}}}

\newcommand\abt{(\ab^{\rm T}_{\rm{eff}})}

\newcommand\afbm{(\ab^\mu_{\rm{eff}})}
\newcommand\cbm{(\cb^\mu)}

\newcommand\afbx[1]{(\ab^{#1}_{\rm{eff}})}
\newcommand\cbx[1]{(\cb^{#1})}

\newcommand\abw{(\ab^w)}
\newcommand\abe{(\ab^e)}
\newcommand\abp{(\ab^p)}
\newcommand\abn{(\ab^n)}
\newcommand\afbe{\afbx{e}}
\newcommand\afbn{\afbx{n}}

\newcommand\cbe{(\cb^e)}
\newcommand\cbp{(\cb^p)}
\newcommand\cbn{(\cb^n)}

\newcommand\cbw{\cbx{w}}
\newcommand\afbw{\afbx{w}}
\newcommand\afw{(a^w_{\rm{eff}})}
\newcommand\cw{(c^w)}

\newcommand\cbpr{(c^{\prime{\rm B}})}
\newcommand\cs{(c^{\rm S})}

\newcommand\cbt{(\cb^{\rm T})}
\newcommand\cbs{(\cb^{\rm S})}
\newcommand\cbb{(\cb^{\rm B})}
\newcommand\cbbp{(\cb^{\prime{\rm B}})}

\newcommand\ctwt{(\ctw{}^{\rm T})}

\newcommand\noe{N_\oplus}

\newcommand\axw{\afbw_{X}}
\newcommand\ayw{\afbw_{Y}}
\newcommand\azw{\afbw_{Z}}
\newcommand\akw{\afbw_{K}}
\newcommand\cxxw{(\cb^w)_{XX}}
\newcommand\cyyw{(\cb^w)_{YY}}
\newcommand\czzw{(\cb^w)_{ZZ}}
\newcommand\cxyw{(\cb^w)_{(XY)}}
\newcommand\cxzw{(\cb^w)_{(XZ)}}
\newcommand\cyzw{(\cb^w)_{(YZ)}}
\newcommand\ctxw{(\cb^w)_{(TX)}}
\newcommand\ctyw{(\cb^w)_{(TY)}}
\newcommand\ctzw{(\cb^w)_{(TZ)}}
\newcommand\cttw{(\cb^w)_{TT}}
\newcommand\ctjw{(\cb^w)_{(TJ)}}

\newcommand\cqb{\cb_Q}
\newcommand\cqbx[1]{(\cb^{#1})_Q}

\newcommand\cxx{\cb_{XX}}
\newcommand\cyy{\cb_{YY}}
\newcommand\czz{\cb_{ZZ}}
\newcommand\cxy{\cb_{(XY)}}
\newcommand\cxz{\cb_{(XZ)}}
\newcommand\cyz{\cb_{(YZ)}}
\newcommand\ctx{\cb_{(TX)}}
\newcommand\cty{\cb_{(TY)}}
\newcommand\ctz{\cb_{(TZ)}}

\newcommand\acc{{\rm a}}

\newcommand\lrpartial{\raise 1pt\hbox{$\stackrel\leftrightarrow\partial$}}
\newcommand\lrprt{\stackrel{\leftrightarrow}{\partial}}
\newcommand\lrprtnu{\stackrel{\leftrightarrow}{\partial^\nu}}
\newcommand\lrDmu{\stackrel{\leftrightarrow}{D_\mu}}
\newcommand\lrDnu{\stackrel{\leftrightarrow}{D^\nu}}
\newcommand\lrvec[1]{\stackrel{\leftrightarrow}{#1} }

\newcommand\atext{$a_\mu$}
\newcommand\btext{$b_\mu$}
\newcommand\ctext{$c_{\mu\nu}$}
\newcommand\dtext{$d_{\mu\nu}$}
\newcommand\etext{$e_\mu$}
\newcommand\ftext{$f_\mu$}
\newcommand\gtext{$g_{\la\mu\nu}$}
\newcommand\Htext{$H_{\mu\nu}$}

\newcommand\eff{{\rm eff}}

\newcommand\G{G_N}

\newcommand\pb{\overline{p}}
\newcommand\nb{\overline{n}}

\newcommand{\beq}{\begin{equation}}
\newcommand{\eeq}{\end{equation}}
\newcommand{\bea}{\begin{eqnarray}}
\newcommand{\eea}{\end{eqnarray}}
\newcommand{\bit}{\begin{itemize}}
\newcommand{\eit}{\end{itemize}}
\newcommand{\rf}[1]{(\ref{#1})}

\newcommand\Qhat{\widehat\Qc}
\newcommand\Shat{\widehat\Sc}
\newcommand\Phat{\widehat\Pc}
\newcommand\Vhat{\widehat\Vc}
\newcommand\Ahat{\widehat\Ac}
\newcommand\That{\widehat\Tc}
\newcommand\Tdual{\widetilde{\widehat\Tc}\phantom{}}
\newcommand\Chat{\widehat\Cc}
\newcommand\Ghat{\widehat\Gc}
\newcommand\Gahat{\widehat\Ga}
\newcommand\Mhat{\widehat M}

\newcommand\mhat{\widehat m}
\newcommand\mfivehat{\widehat m_5}
\newcommand\ahat{\widehat a}
\newcommand\bhat{\widehat b}
\newcommand\chat{\widehat c}
\newcommand\dhat{\widehat d}
\newcommand\ehat{\widehat e}
\newcommand\fhat{\widehat f}
\newcommand\ghat{\widehat g}
\newcommand\Hhat{\widehat H}

\newcommand\pno[1]{PNO(#1)}

\newcommand\nwr[1]{n^w_{#1}}

\newcommand\mm{muonium}
\newcommand\mmb{antimuonium}
\newcommand\hm{H$_\mu$}
\newcommand\Qc{\mathcal Q}
\newcommand\Sc{\mathcal S}
\newcommand\Pc{\mathcal P}
\newcommand\Vc{\mathcal V}
\newcommand\Ac{\mathcal A}
\newcommand\Tc{\mathcal T}
\newcommand\Cc{\mathcal C}
\newcommand\Gc{\mathcal G}
\newcommand\Ec{\mathcal E}
\newcommand\Bc{\mathcal B}
\newcommand\Kc{\mathcal K}
\newcommand\lsim{\mathrel{\rlap{\lower4pt\hbox{\hskip1pt$\sim$}}
    \raise1pt\hbox{$<$}}}

\section{Introduction}
Interest in testing Lorentz symmetry
has blossomed over the past several decades
\cite{proc}.
The renewed interest in the subject
is in part due to the realization
that violations of this symmetry
can be associated with proposed theories of Planck-scale physics
such as string theory \cite{ksp}.
Thus testing Lorentz symmetry
provides a feasible method of gaining experimental information
about Planck-scale physics
with existing technology,
while directly probing the Planck scale
with Planck-energy experiments remains infeasible.

The Standard-Model Extension (SME) is an effective field theory
that incorporates known physics
along with arbitrary, coordinate-independent Lorentz violation
\cite{sme,akgrav,nonminga,nonmin,nonminf}.
In the context of realistic field theory of known interactions,
CPT violation
comes with Lorentz violation \cite{cpt}.
Hence CPT violation is also included in the SME,
which forms a framework for testing these symmetries.
Many phenomena across a wide spectrum of physical systems
have been considered in the context of the SME \cite{jtrev}
and over 1000 experimental and observational sensitivities
to SME coefficients for CPT and Lorentz violation have been achieved \cite{tables}.
Still,
considerable room for additional exploration remains.

The first scientific section of this proceedings contribution
provides an overview of the structure of the SME framework
at the level of the Lagrange density
with special focus on introducing the relatively newer ideas
and notation used in developing the full series of CPT and Lorentz violating
operators beyond the minimal case
of dimension 3 and 4 operators in the fermion sector
without gravitational couplings \cite{nonminf}.
The introduction to nonminimal SME notation continues in Sec.\ \ref{ham}
where we introduce the idea of spherical coefficients for Lorentz violation
in the context of the corresponding hamiltonian.
Section \ref{ngt} reviews recent experimental proposals based on this work
that are of special interest to the antimatter physics community.
In Section \ref{grav}
we introduce gravitational couplings and review their relevant phenomenology.

\section{The SME Expansion}
\label{expansion}

The power of the SME arises in part from the fact
that it is an expansion about the action
for the Standard Model and General Relativity.
As such,
it not only contains known physics,
but it is an action-based theory,
making detailed predictions possible.
Adding CPT and Lorentz violating operators
to known physics
generates an infinite series.
Though such a framework might seem daunting,
methods \cite{nonminga,nonmin,nonminf} have been found that make consideration
of the full series surprisingly tractable,
though some notation must first be absorbed.
The aim of this section is to clarify the relevant notation.
Readers interested in additional details
should consult Ref.\ \cite{nonminf} directly.

An initial simplest form of the Lagrange density for the fermion sector
of the SME (without gravitational couplings) can be written
\beq
\cl = 
\half \psb (\ga^\mu i\prt_\mu - \m + \Qhat) \ps 
+ {\rm h.c.}.
\eeq
Here $\ps$ is the fermion field,
$\m$ is the fermion mass,
and $\Qhat$
contains the series of Lorentz violating terms
involving both $4\times4$ spin matrix operators
and derivatives,
which has been added to known physics
in accord with the SME structure noted above.
In what follows,
the contributions to $\Qhat$
are decomposed in a variety of ways that highlight
features useful in developing experimental sensitivities.

An initial natural decomposition
is to expand $\Qhat$ in the basis of 
Dirac gamma matrices
\beq
\Qhat =
\Shat
+i\Phat \ga_5
+\Vhat^\mu \ga_\mu
+\Ahat^\mu \ga_5\ga_\mu
+\half \That^\mn \si_\mn.
\label{qhatsplit}
\eeq
This makes the $4\times4$ spin-matrix operator structure
explicit, leaving only the momentum operator structure
contained within the Dirac scalars
$\Shat,\Phat,\Vhat^\mu,\Ahat^\mu,\That^\mn$.
This approach will be familiar to those accustomed to the minimal SME,
though the structure might be surprising
as it is more compact.
The point is that we have not yet expanded the derivative operator structure.
In the minimal SME limit,
$\Vhat^\mu = a^\mu + c^\mn p_\nu$
with $p_\mu = i \prt_\mu$.
Here $a^\mu$ and $c^\mn$ are minimal SME coefficients
associated with dimension 3 and 4 operators respectively.

A brief digression from our discussion of SME notation
to remind the reader of the meaning of the terms ``mass dimension'' of operators
and ``CPT properties'' of operators seems useful.
Since the integral of the Lagrange density over $d^4x$ yields the action,
a unitless quantity in natural units,
each term in the Lagrange density must have dimensions of GeV$^4$.
The operator associated with $c^\mn$, $\gamma_\mu i\prt_\nu$,
is referred to as a dimension 4 operator since (other than index structure)
it is the same as the usual kinetic term in the Lagrange density
with $\prt_\mu$ have dimensions of GeV and $\psi$ having dimensions GeV$^{3/2}$.
That makes the associated coefficient $c^\mn$ dimensionless.
The coefficient $a^\mu$ is associated with an operator having no derivatives,
this makes the operator dimension 3 and gives $a^\mu$ units of GeV.
The CPT properties of a given operator can be identified simply
by noting that operators with an even number of free indices are even under a CPT transformation
while operators with odd numbers of indices are odd under CPT.
As background fields,
the coefficients for Lorentz violation do not change under a CPT transformation
of the system (independent of the number of indices they carry).
For an example of how to think about symmetry violation
and background fields,
see Refs.\ \cite{jtrev,berschinger}.

It can also be noted that the simplest expansion \rf{qhatsplit}
in the gamma matrix basis does not perform any separation of the operators
by mass dimension or CPT properties.
The CPT  properties of the operators in the full expansion
can be highlighted using the alternative expansion
\beq
\cl = 
\half \psb (\widehat\Ga^\nu p_\nu - \widehat M) \ps,
+ {\rm h.c.}.
\eeq
where
\bea
\widehat\Ga^\nu &=& 
\ga^\nu
+\chat^{\mn} \ga_\mu
+\dhat^{\mn} \ga_5\ga_\mu
+\ehat^\nu
+i\fhat^\nu \ga_5
+\half \ghat^{\ka\la\nu} \si_{\ka\la} ,
\nonumber\\
\widehat M &=& 
\m 
+ \mhat
+i\mfivehat \ga_5
+\ahat^\mu \ga_\mu
+\bhat^\mu \ga_5\ga_\mu
+\half \Hhat^\mn \si_\mn,
\label{GaM}
\eea
which makes an analogy with the notation used in the minimal SME.
In this expansion,
the hatted letters $a-H$ are associated with operators having even and odd numbers of indices
and are associated with operators that are CPT even and odd respectively.
It should be emphasized that though the analogy is useful here
in identifying CPT properties,
the hatted letters are operators here containing an infinite series
in powers of momentum.
They reduce directly to their minimal SME counterparts
only when the restriction to operators of mass dimension 3 and 4 is made.
For example,
the full content of the operator $\dhat^\mn$ can be written,
\beq
\dhat^\mn p_\nu = \sum_{\rm even} d^{(d)\mu \al_1 \ldots \al_{d-3}} p_{\al_1} \ldots p_{\al_{d-3}},
\eeq
in terms of Cartesian momentum and coefficients
for Lorentz violation at each mass dimension $d$.
The short and $\dhat^\mu = \dhat^\mn p_\nu$ is also sometimes used,
though this complicates the identification of CPT properties.
One can see the minimal SME limit explicitly here
where $\dhat^\mn \rightarrow d^{(4)\mn} = d^\mn$.
Another aspect of the SME's generality is in allowing for the possibility
of different coefficient for Lorentz violation
for different particle species.
Hence protons, electrons, muons, etc.,
each have their own $d^\mn$ coefficients,
for example.
Note also the added possibility of operators $\mhat$ and $\mfivehat$
having no dimension 3 and 4 content.

\section{Spherical Decomposition}
\label{ham}

For most of the experiments of interest below,
the $2\times 2$ hamiltonian $h$ to be acted on 2-component spinors
is the necessary tool to generate a prediction for the impact of Lorentz violation
on the system,
which we'll later expand in powers of $\boldsymbol p/\m$ 
to form the nonrelativistic hamiltonian.
In most cases this amounts to using perturbation theory to find the energy-level shifts.
As in the minimal SME,
generating $h$
from the Lagrange density above proceeds via the standard path.
The only additional challenge in the Lorentz violating case
is associated with obtaining a hermitian hamiltonian in the presence of warping of the time direction,
a challenge which also arises when the Dirac theory is coupled to General Relativity \cite{lvgap,parker}.
This can be handled in the nonminimal case
as in the minimal case using methods that have become standard \cite{nonminf}.

As a simple example of the result,
we consider the hamiltonian associated with $\ahat^\mu$.
As in the minimal SME,
it turns out that not all coefficients are independently observable.
In the case of $\ahat^\mu$, it always appears along with $\ehat$
in the combination $\ahat_\eff^\mu = 
\big( \ahat^\mu - \tfrac{1}{\m} p^\mu \ehat \big)$.
The contribution to the perturbative hamiltonian associated with $\ahat_\eff^\mu$
takes the form 
\beq
h_a = \fr{1}{E_0} \ahat_\eff^\nu p_\nu,
\label{hacart}
\eeq
where $E_0$ is the energy of the unperturbed system.
Expressing $\ahat_\eff^\mu$ in terms of the coefficients it contains
\beq
\ahat_\eff^\mu = \sum_d a_\eff^{(d)\mu\al_1\ldots\al_{d-3}}p_{\al_1}\ldots p_{\al_{d-3}} ,
\label{acart}
\eeq
reveals that analyzing experiments
and reporting sensitivities would become unwieldy at higher $d$
as the number of indices grows.

To address this challenge,
spherical coefficients for Lorentz violation were introduced \cite{nonminga}
and applied to the fermion sector \cite{nonminf}.
The idea is to perform a spherical harmonic decomposition
of relevant quantities.
In the present context,
the decomposition is performed on the perturbative hamiltonian.
In our simple example,
$h_a$ is a rotation scalar and can be decomposed using the usual spherical harmonics as 
\beq
h_a =
\sum_{dnjm} E_0^{d-3-n} |\boldsymbol p|^n\,
\syjm{0}{jm}(\boldsymbol{\hat p})\, \acoef{d}{njm}.
\label{has}
\eeq
The index $d$ in this sum is the dimension of the associated operators.
Since the $a^\mu$ coefficient in the minimal SME is associated with a dimension 3 operator,
$d \geq 3$ here,
and $d$ is odd since the even dimension operators of this type are associated with $\chat^\mn$.
Similarly,
when $d=3$ in the minimal case,
inspection of \rf{hacart} and \rf{acart} reveals
that at most one power of $\boldsymbol p$ should be involved in $h_a$.
Hence $0 \leq n \leq d-2$ in this case.
Since this contribution is spin independent,
the range of values of $j$ can be thought of as being associated with the rotation properties
of a symmetric rank $n$ tensor in the 3 spacial dimensions,
this coming from the powers of 3-momentum that would exist in \rf{acart}
at a given $n$.
Such an object can be decomposed into a series of symmetric traceless tensors
of rank $n,n-2,n-4,\ldots \geq 0$,
each of which corresponds to a value of $j$.
For this example,
this is sufficient to understand the notation for the spherical coefficient
$\acoef{d}{njm}$.
In some cases it is also necessary to specify the parity type as $E$ or $B$.
In the current example,
we see that the parity of the operator is simply set by the spherical harmonic as $(-1)^j$,
which is $E$-type parity ($B$-type is $(-1)^{j+1}$).
For cases in which the hamiltonian is not a rotation scalar
the expansion takes place in terms of the spin-weighted spherical harmonics \cite{nonminga}.
Here the spherical coefficients have extra notation to specify the 
spin weight and parity type.
For example,
$g_{njm}^{(d)(1E)}$
is a spherical coefficient associated with $E$-type parity and spin-weight 1 (number preceding the $E$).
The spin-weight is given by the helicity of the operator
with spin-weight 1 appearing for spin 1/2 fermions,
spin-weight 2 appearing for photons, etc.

Use of spherical coefficients generates a set of quantities that essentially get no more complex
as one goes to higher mass dimension operators.
It is also the case that they transform easily under rotations,
an advantage since many Lorentz-violation searches
involve testing rotation invariance
and rely on transformations between rotated frames.
The relation between the spherical coefficients
and the Cartesian ones
can be easily demonstrating by comparing the spherical expansion for the minimal case
with the Cartesian minimal hamiltonian.
One finds,
for example,
that 
\beq
\acoef{3}{110} = -\sqrt{\fr{4\pi}{3}} a^{(3)z}_\eff,
\eeq
where in the notation of the minimal SME $a^{(3)z} \rightarrow a^z$.

In the systems to follow,
the nonrelativistic limit of $h$ is of interest.
Here the usual expansion
\beq
E_0 \approx \m + \fr{|\boldsymbol p|^2}{2\m} - \fr{|\boldsymbol p|^4}{8\m^3} +\ldots.
\label{nreomz}
\eeq
for the energy can be inserted into $h$.
In the example of $h_a$
the result can be written
\beq
h_a = \sum_{njm} |\boldsymbol p|^n \, \syjm{0}{jm}(\boldsymbol{\hat p}) 
\bigg(\sum_d \m^{d-3-n}
\sum_{k\leq n/2} \bc{(d-3-n+2k)/2}{k} \acoef{d}{(n-2k)jm} \bigg),
\qquad 
\label{haexp}
\eeq
where $\bc{j}{k}$ is a binomial coefficient.
The quantity in parenthesis is then given the name $a_{njm}^{\rm NR}$,
or the nonrelativistic spherical coefficients for Lorentz violation,
so named since they denote the coefficients which govern Lorentz-violating
effects with a given $jm$ and power of $|\boldsymbol p|$ in nonrelativistic tests.
Note that the nonrelativistic expansion here
mixes coefficients at different $d$.
For example,
\beq
a_{100}^{\rm NR} = \m^{-1} \acoef{3}{100} + \m \acoef{5}{100} + \m^{3} \acoef{7}{100} + \ldots
\eeq

\section{Nongravitational Tests}
\label{ngt}

Within the past 2 years,
a pair works have been published that significantly extend the potential for SME-related studies
in systems of interest to the LEAP community.
Most recent is  Ref.\ \cite{nmlvH},
which develops experimental proposals to search for higher dimension Lorentz violation
with hydrogen, antihydrogen, and related systems 
over the course of 37 pages in Physical Review D.
The discussion provided extends the minimal work
on these systems \cite{minH}.
Reference \cite{akmu} does the same for muons and related systems
in a span of 24 Physical Review D pages,
building on the prior discussion of these systems that was restricted to the minimal case \cite{smemu}.
After some discussion of how the general theory reviewed above
is to be applied in the relevant systems,
both works develop specific predictions for a wide variety of experiments,
in many cases providing specific formula to which experimental data can be fit.
Given this vast body of useful material,
the goal of this section is to summarize the content of these 61 pages such that the interested researcher
might be made efficiently aware of the potential
that systems available to them might have.

\subsection{Hydrogen and Antihydrogen}

The discussion of relevant experimental systems in Ref.\ \cite{nmlvH}
begins in Sec.\ III with the discussion of spectroscopic studies in hydrogen.
The first of 3 subsections addresses signals in free hydrogen
without applied electric and magnetic fields.
In this context,
relevant Lorentz-violating effects can be explored via
a Lorentz-violating splitting of otherwise degenerate levels,
which is analogous to the hyperfine Zeeman splitting.
Here the levels can be split by coefficients for Lorentz violation
that come together to form a vector $\mathbf A$
that plays a role analogous to the role of the magnetic field
in the Zeeman effect.
If one imagines that transitions among these levels were
excited by a laser of fixed polarization in the lab,
the variations of in the vector $\mathbf A$
induced by the sidereal rotation of the lab
would lead to variations in the transition probabilities
at harmonics of the sidereal frequency.
A unique feature of this effect is that the transition probabilities
are an unsupressed Lorentz-violating effect
proportional to a ratio of coefficients for Lorentz violation;
however their observation hinges on the ability to resolve
the potential Lorentz-violating energy splitting,
an effect that is suppressed.
The above effect is accompanied by sidereal variations in the observed line shapes.
Experimental options are noted along with the entertaining insight
that if Lorentz-violating splittings were observed,
they could be used to create a novel kind of maser.

The analysis of spectroscopic studies in hydrogen
provided by Ref.\ \cite{nmlvH} continues
in the second subsection of Sec.\ III
with consideration of perturbative effects
on the conventional hyperfine Zeeman transitions
associated with weak magnetic fields.
Prospects for investigations of rotation-invariance violation
that make use of
the sidereal rotation of the laboratory
are considered as are
searches that could be done by rotating the system in the laboratory.
Predictions associated with boost-invariance violation
that can be studied using annual variations
are developed,
and finally versions of these effects
with space-based experiments are described
that make used of the additional rotations and boosts
available on a space-based platform.

The penultimate subsection of Sec.\ III
considers transitions with $J=1/2$ and $\Delta J =0$
with focus on $nS_{1/2}-n^\prime S_{1/2}$
and $nS_{1/2}-n^\prime P_{1/2}$,
including the popular $1S-2S$ transition.
The features of the nonminimal coefficients
have several implications for the study of these transitions.
Since the relevant coefficients have mass dimension,
the absolute sensitivity will be a more important factor
in assessing sensitivity to CPT and Lorentz violation
than the relative precision.
Since the attainable sensitivity is lower here,
spin-dependent effects,
which can be better measured in the hyperfine Zeeman transitions
are not included in the analysis.
Finally,
those familiar with SME history will recall that 
these transitions lacked leading sensitivities
to minimal SME coefficients.
Reference \cite{nmlvH} demonstrates
that unlike the minimal case,
leading sensitivity to nonminimal coefficients indeed arises,
including sensitivity to isotropic coefficients
that are not readily accessible in Zeeman transitions.
The final subsection of Sec.\ III considers transitions
involving a level with $J$ or $F$ greater than or equal to $3/2$.
A special feature of these transitions is the appearance of
spectral frequency variations at the 3rd and 4th harmonic
of the sidereal frequency.

Following discussion of hydrogen,
the paper considers most of the same opportunities
in antihydrogen.
Although the same coefficients for Lorentz violation
are involved in the relevant signals in antihydrogen,
different linear combinations appear due to the fact
that the relevant combinations involve both CPT odd and CPT even
contributions.
Though these contributions could in principle be separated
in highly boosted systems without the use of antimatter,
this is very likely impractical in practice.
The isotropic invisible model \cite{lvgap}
illustrates the point via a minimal example in a spectroscopic context,
while the same idea is illustrated by the isotropic parachute model (IPM)
discussed in some detail in Sec.\ \ref{grav}.
Access to these coefficients will be much more readily available
to developing precision antihydrogen spectroscopy experiments \cite{hbar}.

The remaining sections of the paper address 
possibilities for measurements of Lorentz violation
in various related systems.
Deuterium spectroscopy,
comparisons of the Rydberg constant measured in Deuterium
verses measurements in hydrogen,
and measurements with a Deuterium maser are considered.
The Deuterium maser is a particularly interesting possibility
as it could offer many orders of magnitude improvement
over hydrogen-maser measurements.
Finally,
the potential of spectroscopic studies of positronium and
of hydrogen molecules is developed.

\subsection{Muonic systems}

The proposals in Ref.\ \cite{akmu}
probe independent effects from those of Ref.\ \cite{nmlvH}
due to the particle-species dependence of the SME coefficients noted in Sec.\ \ref{expansion}. 
Here the focus is clearly on the muon coefficients.
The exploration of relevant muonic systems begins in Sec.\ II
of Ref.\ \cite{akmu} with consideration of spectroscopic measurements in muonic bound states.
Predictions for SME effects on transitions in \mm\
are provided in Sec.\ IIB.
This includes hyperfine transitions,
the $1S$-$2S$ transition, and the Lamb shift.
The explicit frequency shift due to the relevant SME coefficients
is provided for each of these cases respectively by Eqs.\ (8), (12), and (14)
of Ref.\ \cite{akmu}.
Initial constraints are placed on some SME coefficients
using these equations
along with the published results of hyperfine transition measurements \cite{hugh}.
Planned work \cite{shim} could lead to 5-fold improvement
on these sensitivities.
The complementary set of isotropic coefficients
are constrained via comparisons of existing experimental values for the 
$1S$-$2S$ transition and the Lamb shift with the theoretical values.
Section IIC 
provides an analysis of SME effects in muonic atoms and ions.
The possibility future searches
for sidereal variations in muonic hydrogen (\hm) Zeeman transitions
is explored.
The frequency shift 
of the $2S^{F-1}_{1/2}$-$2P^F_{3/2}$ transitions
induced by Lorentz violation
is provided by Equation (18) of Ref.\ \cite{akmu},
and it is shown that interesting sensitivities in future experiments
can be attained.

The so-called `proton radius puzzle' \cite{pohl}
is an apparent disagreement that currently exists between
proton charge radius measurements obtained 
from \hm\ spectroscopy and from other types of tests.
It is shown in Sec. IIB   
that certain SME coefficients,
if found to have
suitable nonzero values,
would generate a frequency shift in \hm\
that would ``explain'' the observed disagreement.
The values for SME coefficients required are permitted by current constraints.
The discussion of \hm\ spectroscopy concludes with
consideration of a method useful in searching for SME coefficients
when the Zeeman transitions are unresolved,
which involves apparent broadening of the spectral lines
due to rotational symmetry breaking.
Section IIB  concludes
with a discussion of the prospects for studying Lorentz and CPT violation 
using other muonic atoms and ions.

In Sec.\ III of Ref.\ \cite{akmu} searches for Lorentz and CPT violation
using anomalous magnetic moment measurements 
of the muon and antimuon are addressed.
Following some theoretical analysis to obtain
the shift in the anomaly frequency due to the full SME expansion,
which is provided by Ref.\ \cite{akmu} Eq.\ (42),
methods of obtaining interesting sensitivities to SME coefficients are developed.
Some initial constraints are then placed using existing results
from 
Brookhaven National Laboratory (BNL) \cite{bnl}
and CERN \cite{cern},
on which 5 fold improvements are possible in
upcoming experiments
at the Japan Proton Accelerator
Research Complex (J-PARC) \cite{j-parc} and at the Fermi National Accelerator Laboratory (Fermilab)
\cite{fermilab}.
The first method discussed involves comparison of the muon and antimuon 
anomaly frequencies
using different schemes to separate constraints
on CPT-odd and even coefficients.
The potential for use of periodic variations in the anomaly frequency
at characteristic frequencies associated with Lorentz violation
(sidereal and harmonics there of, and annual)
is then highlighted.

Final comments on anomalous magnetic moments
focus on the  apparent disagreement known as the
`anomaly discrepancy'.
Here the disagreement is between calculations of the muon anomaly performed
within the SM \cite{jn} and the BNL results \cite{bnl,blum}.
The paper again notes that certain nonzero SME coefficients
could generate such a discrepancy
with coefficient values that are consistent
with existing limits.

Clearly the developments in Ref.\ \cite{akmu} amount to a considerable expansion
of proposals to search for Lorentz violation in the muon sector;
however discussion concludes with notes on further possibilities for expansion.
Ideas listed include
additional techniques for measuring the SME coefficients considered here,
the possibility of including interactions in the analysis,
the inclusion of nonminimal gravitational couplings,
consideration of flavor-changing effects involving muons,
and the prospects for performing similar analysis with other particles.

\section{Gravitational Couplings}
\label{grav}

The effects of Lorentz violation in gravitational tests
can originate from the pure-gravity sector \cite{lvpn,gexpt,solarsys},
including some explorations of higher-order operators \cite{nmgrav},
or from gravitational couplings in the other sectors \cite{lvgap,akjt}.
In section \ref{expansion}
we introduced the idea of the SME expansion
in the context of the fermion sector in flat spacetime,
and considered the full series of Lorentz violating operators in that context.
Here we introduce the fermion sector of the SME
with gravitational couplings for case of dimension 3 and 4 operators,
sometimes referred to as the gravitationally coupled minimal fermion sector \cite{akgrav},
and discuss its implications:
\beq
\cL_{\ps-g} = 
\half i e \ivb \mu a \ol \ps \Ga^a \lrDmu \ps 
- e \ol \ps M \ps,
\label{fermion}
\eeq
where
\bea
\Ga^a
&\equiv & 
\ga^a - c_{\mu\nu} \uvb \nu a \ivb \mu b \ga^b
- d_{\mu\nu} \uvb \nu a \ivb \mu b \ga_5 \ga^b
- e_\mu \uvb \mu a 
- i f_\mu \uvb \mu a \ga_5 
- \half g_{\la\mu\nu} \uvb \nu a \ivb \la b \ivb \mu c \si^{bc}, \\
M
&\equiv &
m + a_\mu \ivb \mu a \ga^a 
+ b_\mu \ivb \mu a \ga_5 \ga^a 
+ \half H_{\mu\nu} \ivb \mu a \ivb \nu b \si^{ab}.
\eea
The notation $\cL_{\ps-g}$ is used here to distinguish this limit of the fermion sector
from that of Sec.\ \ref{expansion}.
Despite the similarity in the symbols used,
\atext, \btext, \ctext, \dtext, \etext, \ftext, \gtext, \Htext\
here (no hats) 
are coefficient fields for Lorentz violation,
a concept we develop further below,
which is distinct both from the operators introduced above
and the coefficients for Lorentz violation used in the minimal SME without gravitational couplings,
though there is a connection to the latter.
The object $\vb \mu a$ is the veirbein,
which provides the gravitational couplings
by linking each point on the spacetime manifold with a Minkowski tangent space.
The determinant of the veirbein, denoted $e$,
and the covariant derivative
also generate gravitation contributions in $\cL_{\ps-g}$.
Note that the replacement $\vb \mu a \rightarrow \de^a_\mu$
along with the appropriate interpretation of the covariant derivative
and the coefficients for Lorentz violation
leads to the form of the Minkowski-spacetime minimal fermion-sector.

The coefficients for Lorentz violation
in the Minkowski spacetime SME are typically taken as constant (i.e.\ $\prt_\al c_\mn=0$).
In this limit the coefficients can be thought of as existing in this form 
either as an externally prescribed property of the spacetime (explicit Lorentz violation)
or as the vacuum values associated with spontaneous breaking of Lorentz symmetry
(spontaneous Lorentz violation) \cite{akrb},
the latter involving a process that can be thought of by analogy with $SU(2) \times U(1)$
breaking in the Standard Model.
While the distinction between the 2 scenarios for how Lorentz violation might arise
is not important in nongravitational experiments in Minkowski spacetime
seeking effects associated with the vacuum values,
explicit Lorentz violation
is incompatible with most gravity theories based on Riemann geometry \cite{akgrav,rb}.
Hence consideration is specialized to the case of spontaneous Lorentz violation
in the current section,
where a consistent treatment requires the consideration of the fluctuations
about the vacuum values in addition to the vacuum values themselves.
After spontaneous symmetry breaking,
the coefficient fields of \rf{fermion}
can be written, for example,
\beq
c_\mn = \cb_\mn + \ctw_\mn,
\eeq
where the 2 terms here respectively denote the vacuum values (taken as constant)
and the fluctuations.
In a post-newtonian analysis the fluctuations
can typically be eliminated in favor of the vacuum values
and the gravitational field.
Observables can then be expressed in terms of the vacuum values,
which can be identified with those of the Minkowski-spacetime SME
(where they appeared without the bar).

A detailed analysis of the post-newtonian experimental and observational
implications of the spin-independent coefficient fields
\atext, \ctext,\ and \etext\ is provided by Ref.\ \cite{lvgap}.
In addition to the vacuum value associated with \ctext\ introduced above,
the vacuum value $\afb_\mu = \ab_\mu - m \eb_\mu$ is the other associated
coefficient for Lorentz violation.
This coefficient is of special interest in gravitational studies
as it is known as a countershaded combination,
a coefficient that is unobservable except in special circumstances 
such as via gravitational experiments \cite{akjt}.
Implications of these coefficients,
particularly those relevant for LEAP-related systems are reviewed here.
Spin couplings are also of interest \cite{gspin},
though our current focus will remain on the spin-independent coefficients above.

Gravitational tests that are relevant in the search for Lorentz violation
in the fermion sector of the SME include \cite{lvgap}
solar-system tests \cite{solarsys},
experiments with devices traditionally used for short-range gravity tests \cite{db},
spin-precession tests \cite{jtspin},
tests of the universality of free fall \cite{bke,ste},
and
redshift tests \cite{bke,red}.
The key ideas can be naturally understood
in the case of laboratory tests
near the surface of the Earth
involving Earth's field. 
The coefficients for Lorentz violation generate tiny corrections
to the gravitational force
both in the direction along the usual free-fall direction
and perpendicular to it.
Certain coefficients also have the practical effect of generating a
direction-dependent inertial mass.
This
results in a nontrivial relation between force and acceleration \cite{berschinger}.
As the Earth-lab system boosts and rotates,
these effects are time dependent having variations
at the annual and sidereal frequencies,
as is typical in Lorentz violation studies,
and
may also be particle-species dependent
due to the generality of the SME in allowing particle-species dependent coefficients
for Lorentz violation.
As a result of the properties summarized above,
lab tests were divided into 4 categories:
(i) a free-fall gravimeter tests,
which measure the gravitational acceleration of a body in free fall
as a function of time,
(ii) force-comparison gravimeter tests,
which measure the gravitational force as a function of time,
(iii) free-fall Weak Equivalence Principle 
(WEP) tests,
which monitor the relative acceleration of a pair of freely falling bodies,
and (iv) force-comparison WEP tests,
which monitor the relative force.

Versions of the above tests performed with novel types of matter
such as antimatter, second- and third-generation particles,
and charged particles
have the potential to attain sensitivities to coefficients for CPT and Lorentz violation
that are challenging or impossible to measure via conventional gravitational tests.
Ref.\ \cite{lvgap} develops predictions for gravitational experiments
with antihydrogen \cite{hbarg,muelcharge},
muonium \cite{muon}
and charged particles \cite{muelcharge,chargeai,charge}.
Other exotic atoms containing antiparticles \cite{positronium} or higher-generation matter
may also be of interest in this context.
Antimatter tests have the potential to contribute to efforts to obtain
independent sensitivities $\afb_\mu$ and $\cb_\mn$ coefficients
because
the sign of $\cb_\mn$ terms does not change under CPT
while in the case of $\afb_\mu$ it does.
For this reason
antimatter experiments
could place cleaner constraints
on certain combinations of SME coefficients
than matter-only tests
and perhaps could observe novel behaviors
stemming from CPT and Lorentz violation in the SME.
The particle species dependence of coefficients implies
unique sensitivity to certain SME coefficients,
such as $\afb_\mu$
for the muon \cite{muon},
are possible in gravitational experiments with higher-generation matter.
There are also combinations of SME coefficients that appear to be observable
only in gravitational experiments with charged matter.
Hence gravitational experiments with charged matter may also have the potential
to reveal comparatively large Lorentz violation in nature.

Numerous arguments against anomalous antimatter gravity
have been developed in the literature
based on indirect constraints \cite{mntg}.
The general field theory based approach of the SME
may also illuminates some aspects of these limits.
By definition,
an indirect constraint must happen within a model.
That is,
the model must be used to predict the results of an experiment
based on the known outcome of others.
Indirect constraints found using one model may not apply to others.
Certain limits of the full SME generate toy models
than can be used to explore indirect constraints on anomalous antimatter gravity.
The isotropic `parachute' model (IPM) \cite{lvgap},
provides one example.
The IPM is developed 
by restricting the classical nonrelativistic Lagrange density of the SME  
to the limit in which 
$\afbw_T$ and isotropic $\cbw_{\Si\Xi}$
are the only nonzero coefficients
in the Sun-centered frame.
In which case,
the effective classical Lagrangian 
for a test particle T moving in the gravitational field
of a source S
takes the suggestive form
\beq
L_{\rm IPM} = \half \mt_i v^2 + \fr{\G \mt_g \ms_g}{r},
\eeq
with $v$ being the the velocity,
$r$ the distance from the source,
$\mt_i$ the effective inertial mass of T,
and $\mt_g$ and $\ms_g$ 
the respective effective gravitational masses 
of bodies T and S,
which take the explicit form:
\bea
\nonumber
m^{\rm B}_i &=& 
m^{\rm B} + \sum_w \fracn53 (N^w+N^{\bar{w}}) m^w \cbw_{TT} \\
m^{\rm B}_g &=& 
m^{\rm B} + \sum_w \Big( (N^w+N^{\bar{w}}) m^w \cbw_{TT}
+ 2 \al (N^w-N^{\bar{w}}) \afbw_T \Big).
\label{friedmasses}
\eea
Here B is T or S
with $m^{\rm B}$ being the conventional (Lorentz invariant) body mass,
$m^w$ the mass of a particle of species $w$,
$N^w$ and $N^{\bar{w}}$
the respective number of particles and antiparticles of species $w$
in B.
The IPM's defining conditions:
\beq
\al \afbw_T = \fracn 13 m^w \cbw_{TT},
\label{defipm}
\eeq
are then imposed to complete the construction of the model.

The result of the IPM conditions
is equal effective gravitational and inertial masses
for a matter body,
$m^{\rm B}_i = m^{\rm B}_g$,
implying that no Lorentz-violating effects 
arise in tests with ordinary matter to post-newtonian order 3.
For antimatter bodies, the situation is different,
$m^{\rm B}_i \neq m^{\rm B}_g$.
Hence the comparison
of the gravitational responses 
of matter and antimatter
or of different types of antimatter
may result in observable signals within the IPM.
Some arguments against
anomalous antimatter gravity 
have been explored in the context of the IPM \cite{lvgap, wag}.
Though many traditional indirect limits appear ineffective
at constraining the IPM,
it can be constrained using experiments involving multiple boost factors \cite{lvgap}
and using higher post-newtonian order studies \cite{bke,red}.
A generalization of the IPM using coefficients from the nonminimal SME has been developed
that remains to be constrained \cite{nmlvH}.

\end{document}